\documentclass[12pt]{article}
\usepackage{epsfig}
\begin{document}

\title{\bf Defining statistical ensembles of random 
graphs\thanks{Talk at the workshop "Discrete random 
geometries and quantum gravity", Utrecht, October 2001}}
\author{Andr\'e Krzywicki \\ Laboratoire de Physique 
Th\'eorique, B\^atiment 210\\
Universit\'e Paris-Sud, 91405 Orsay, 
France\thanks{Unit\'e Mixte du CNRS UMR 8627.}}
\date{ }
\maketitle

\begin{abstract}
The problem of defining a statistical ensemble of random graphs
with an arbitrary connectivity distribution is discussed. 
Introducing such an ensemble is a step towards  
uderstanding the geometry of wide classes of graphs 
independently of any specific model. This research was 
triggered by the recent interest in the so-called scale-free 
networks.
\end{abstract}
\vspace{0.8cm}
LPT Orsay 01/102\\
October 2001
\vspace{0.8cm}
\par\noindent
\section{Introduction}
This is a workshop talk and therefore I do not hesitate 
to report about partial results of a research still 
in progress. I shall also submit you a couple 
of queries, with the hope of attracting your interest and 
triggering a discussion. I have benefited from collaboration 
with Z. Burda, the late J.D. Correia and J. Jurkiewicz (cf 
ref. \cite{bck} and papers quoted therein). 
\par
Let me recall that a graph is just a collection of vertices 
(nodes) and links (edges) connecting vertices. It is a 
mathematical idealization representing 
various networks one encounters in nature, 
in social life, in engineering, etc. 
For example, the web can be represented by a 
graph: the vertices are the URLs 
and the links are the hyperlinks. Likewise, the 
network of sexual relations in 
a population can be represented by a graph. The 
study of its geometry has some 
interest for epidemiology. In these examples, 
as in many other ones, the pattern 
of connections between vertices is fairly random. 
The concept of a {\em random 
graph} emerges quite naturally. For definiteness, 
I shall consider graphs 
with undirected links only.
\par
When one is talking about {\em random graphs}, 
one has of course in mind a 
statistical ensemble of graphs. How to 
define such an ensemble? The simplest 
answer is given in the framework of the 
classical model developed by Erd\"os, R\'enyi and 
their followers \cite{er}: in a set of $N$ 
vertices one connects at random $L$
out of $N(N-1)/2$ possible pairs of vertices. 
All possible graphs constructed 
that way form the ensemble in question. 
The probability $p$ to connect a
pair of vertices is the control parameter 
of the model. The geometry of graphs
changes in a very interesting and by now 
fully understood manner when $p$ 
changes. However, in this ensemble the 
distribution of connectivity (vertex degree) 
is always Poissonian. 
\par
It turns out that connectivity distributions 
very different from Poissonian
are observed in a variety of observed networks. 
In particular, in a number
of interesting networks this distribution has 
a tail falling like a power of 
the vertex degree. These networks have been 
baptized {\em scale-free} by 
Barab\'asi and Albert \cite{ba}. The properties 
of scale-free networks
are commonly discussed in the framework 
of simple growth models (where the
connectivity distribution becomes stationary
and scale-free at large time). These
models are invaluable for illustrating 
basic dynamical mechanisms, like
the preferential attachment rule. However, 
they are not fully realistic.
For a variety of reasons one 
would like to understand the generic
geometries of wide classes 
of graphs. This can be presumably
better achieved by defining consistently 
the corresponding statistical
ensembles, instead of producing more 
and more complicated growing
network models.
\par
The aim of this talk is to discuss 
problems one encounters trying to define 
a statistical ensemble of random 
graphs with an {\em a priori} 
given connectivity distribution. The 
definition can be more or less formal. 
It can be implicit, reducing to the 
formulation of an algorithm enabling one 
to sample graphs, for example 
with the help of a computer.

\section{The Molloy-Reed construction}
Let $p_n$ denote here the connectivity 
distribution. In ref. \cite{mr} 
Molloy and Reed propose a specific method 
of constructing graphs with a given 
$p_n$. They proceed in two steps:
\par
\hspace{0.8cm} (a) First, $N$ auxiliary 
graphs are created. The number of links 
of an auxiliary graph is randomly generated 
from the probability distribution 
$p_n$. By construction all these links meet 
at a common vertex and have the other 
end free. The number of free-end links in 
the full set of auxiliary graphs must 
be even, otherwise one restarts the construction.
\par
\hspace{0.8cm} (b) Second, in the full set 
of $N$ auxiliary graphs the successive 
pairs of free link ends are picked at 
random and connected, until no free link 
end remains. 
\par
In this manner, one creates a single graph 
with vertex degrees $n_1, n_2, ... n_N$.
Notice, that the number of links of that 
graph $L = 1/2 \sum_j n_j$ is not kept
fixed. In the ensemble of graphs 
it does fluctuate around the average
value $1/2 N \langle n \rangle$, where 
$\langle ... \rangle = \sum_n ...\; 
p_n$. This is perhaps a weak point of 
the construction, since $L/N$ is a sensitive
parameter in graph theory. On the other 
hand, it is very pleasant that the 
connectivity distribution matches 
$p_n$ for {\em individual graphs}.
\par
Notice also that these graphs are, in 
general, not connected. Furthermore, they are,
in general, "degenerate": there may be 
multiple connections between vertices and
certain links may connect a vertex to 
itself \footnote{I met the opinion that in
this construction the degenerate graphs 
become unimportant in the limit 
$N \to \infty$. This is false. It 
is easy to count graphs. When 
$N \to \infty$ and $x = L/N$ is 
kept fixed, the non-degenerate graphs 
are a finite fraction, viz. $\exp{[-2x(1+x)]}$, 
of all possible graphs.}. 
For a given set $n_1, n_2, ... n_N$
a non-degenerate graph may simply not 
exist. Moreover, enforcing non-degeneracy,
when it is possible, introduces a bias. 
Although in each graph the connectivity 
distribution matches $p_n$ up to fluctuations, 
significant deviations from $p_n$ 
can appear, when the distribution is 
calculated for a large ensemble of graphs, 
if certain fluctuations are systematically 
favoured. 
This remark is particularly
pertinent to the case of scale-free graphs, 
where the connectivity distribution
has a long tail, subject to important fluctuations. 

\section{Minifield theory: random 
graphs and Feynman diagrams}
The minifield theory is defined by 
the following formal integral
\begin{equation}
Z \sim \int d\phi \exp{\frac{1}{\kappa}[-\phi^2/2\lambda 
+ \sum_n p_n \phi^n]}
\label{5}
\end{equation}
\noindent
where the integration variable $\phi$ is a 
real number, $\kappa, \lambda, p_1 > 0$ 
and $p_n \ge 0$ for $ n>1 $. Although, 
strictly speaking, the integral does not
exist, the perturbative expansion of $Z$ 
in powers of the "couplings" $p_n$ is
well defined. As in field theory, the 
individual terms of the expansion
can be represented by Feynman diagrams. 
The "propagator" equals $\lambda$,
$\kappa$ plays the role of the Planck constant 
and $p_1$ that of an "external
current" (a pedagogical presentation for 
people not very familiar with field theory
methods can be found in \cite{bck}).
\par
The idea is to identify the Feynman 
diagrams of this toy model with the graphs 
of a statistical ensemble. Indeed, the 
Feynman diagrams of the minifield theory 
are the graphs familiar to people working 
on networks, except that there is a 
specific weight - the "Feynman amplitude" - 
attached to each graph. In the 
"semiclassical limit" $\kappa \to 0$ only 
tree graphs survive and the model is
exactly solvable.
\par
According to the Feynman rules, the weight 
of a non-degenerate graph with $N$
vertices and $L$ links is
\begin{equation}
weight = \kappa^{L-N}\; \frac{\lambda^L}{N!}\; 
\prod_{j=1}^N \; [p_{n_j} n_j!]
\label{6}
\end{equation}
\noindent
In the presence of degeneracies one has 
to multiply the rhs by the standard
symmetry factors. Actually, the construction 
of Feynman diagrams does not differ
from the construction of graphs following 
the Molloy-Reed recipe. Here, the 
auxiliary graphs are those defined by the 
"interactions" $p_n \phi^n$. However, 
the weight factor $\kappa^{L-N} \lambda^L/N!$ 
does not appear there; 
the fluctuations of $L$ result from 
fluctuations of the generated vertex
degrees. In contrast, we introduce here a 
specific fugacity of links $\lambda$ and
a parameter, $\kappa$, controlling the
 number of loops in connected components.
\par
The following Metropolis algorithm generates 
graphs with fixed $N$ and $L$: one 
picks a random link $\vec{ij}$ and a random 
vertex $k \ne i,j$ and one rewires 
$\vec{ij} \to \vec{ik}$ with probability
\begin{equation}
P_{rewire} = (n_k+1) R(n_k+1)/n_j R(n_j)
\label{7}
\end{equation}
\noindent
when the rhs above is less than unity, and
with probability equal to one otherwise.
Here $R(n) = p_n/p_{n-1}$. When $n_j=1$, 
the attempt is rejected, so that vertices
with zero connectivity are never created. The 
rhs of (\ref{7}) follows from  
(\ref{6}) and the detailed balance condition. It 
turns out, that this last condition
insures that the symmetry factors in the weights 
of degenerate graphs come out 
correctly too.
\par
The presence of the factor $(n_k+1)/n_j$ on 
the rhs of (\ref{7}) means that the
rewired vertices are sampled independently 
of their degree. Furthermore, the
rewiring depends on the vertex degrees only 
and is insensitive to the rest of the
underlying graph structure. Hence, as far as 
the distribution of vertex degrees is
concerned, the model is isomorphic to the 
well known balls-in-boxes model 
\cite{binb}, defined by the partition function
\begin{equation}
z \sim \sum_{n_j} p(n_1) ... p(n_N) \delta(M - 
\sum_{j=1}^N n_j)
\label{8}
\end{equation}
\noindent
and describing $M$ balls distributed with 
probability $p_n$ among $N$ boxes
(in our case $M=2L$). The constraint 
represented by the Kronecker delta on the
rhs of (\ref{8}) is satisfied "for free" 
when $N \to \infty$ by virtue of 
Khintchin's law of large numbers, provided 
$\langle n \rangle < \infty$ and 
$M/N = \langle n \rangle$. When the
last condition is met the occupation 
number distribution of a single 
box is just $p_n$.
\par
Consequently, in the statistical ensemble
including {\em degenerate} graphs 
the connectivity distribution is 
$p_n$ provided the number
of links is set to $L = 1/2 N \langle n \rangle$ 
(notice, that it is the
average number of links in the Molloy-Reed 
construction). It is easy to calculate 
the number of such graphs for fixed $L/N$ . 
It increases with $N$
like $\exp{[\mbox{\rm const} N \log{N}]}$, 
the ensemble is overextensive~\footnote{
The ensemble of non-degenerate graphs is 
overextensive too (cf the footnote 
on p. 3); it becomes extensive
in the limit $\kappa \to 0$, ie for tree 
graphs.}. Hence, it is not guaranteed that
the connectivity distribution is $p_n$ for 
individual graphs, it is so when one
averages over the ensemble. This should 
not be a serious flaw in applications.
\par
The algorithm works also very well for {\em trees}. 
It suffices to start with a tree
graph, for example with a polyline, and impose 
the constraint that $n_i=1$. Then,
all successively generated graphs are also trees. 
As already mentioned, the model
is analytically solvable when one limits one's 
attention to tree graphs. One can
show exactly that in this case the connectivity 
distribution is $\sim np_n$ \footnote{The 
following heuristic argument can help to understand
that: trees can always be embedded
in a plane. They are obtained by gluing
successive vertices (auxiliary graphs). But each vertex 
with $n$ links attached to it could have been rotated
in the plane up to $n$ times before beeing
glued to the tree it belongs to and
this rotation would not affect the result.
Consequently, the weight of a vertex
is $\sim np_n$ instead of $p_n$,
because of this specific symmetry.}. Hence, in
order to get an {\em a priori} given connectivity 
distribution $P_n$ one should
set the couplings of the tree model to $p_n \sim P_n/n$. 
\par
A fairly comprehensive discussion of the ensemble 
of random tree graphs is
presented in \cite{bck}, with emphasis on the 
hot problem of scale-free 
graphs. I shall not enter into this discussion 
here, apart 
from the few words to follow. The partition 
function (\ref{5}) can be calculated 
in the saddle point ("semi-classical") 
approximation. The saddle point condition, 
identical to a familiar equation in 
polymer physics, is a starting point for further
calculations. In particular, one can 
find the fractal dimension $d_H$ of 
the tree graphs. This was first done 
in \cite{adj} for the so-called generic 
case, with the result $d_H=2$. For 
scale-free graphs the 
connectivity distribution falls 
like $n^{-\beta}$ and one finds \cite{jk,bck} 
in the empirically interesting 
situation $2 < \beta \le 3$:
\begin{equation}
d_H = (\beta-1)/(\beta-2)
\label{9}
\end{equation}
\noindent 
while $d_H=2$ again for $\beta > 3$. An 
infinite $d_H$ is found in the rather
special case, where a singular vertex 
with fixed degree of order $O(N)$ is
present in (almost) all trees of the 
ensemble \footnote{The Cayley tree,
a graph with a minimal entropy 
in our ensemble, also has $d_H = \infty$.}.
\par
It is very easy to supplement the algoritm 
with a constraint insuring that all
produced graphs are non-degenerate. However, 
this introduces a bias. We do not know yet how 
to choose the input data, ie the couplings 
$p_n$, in order to get at the output 
a desired connectivity 
distribution. The problem is solved 
for degenerate graphs and for trees, 
as stated above, but for non-degenerate graphs
it remains open:
\par
{\em Query} : What are the minifield 
theory couplings $p_n$ leading to a
given connectivity distribution in the 
ensemble of {\em non-degenerate} graphs ? 
\par
The ensemble of graphs defined by (\ref{5}) 
is fairly general, but not the most
general one: the weight of a graph is a 
product of factors corresponding to
individual vertices. One can introduce 
correlations between neighbor vertices
replacing (\ref{5}) by
\begin{equation}
Z \sim \int d^q\phi \exp{\frac{1}{\kappa}[-\vec{\phi} 
A\vec{\phi} + \sum_{n=1}^q \phi_n^n]}
\label{10}
\end{equation}
\noindent
where $\vec{\phi} = (\phi_1, \phi_2, ..., \phi_q)$ 
and $A$ is some $q \times q$ symmetric 
matrix with positive elements. The cut-off
$q$ can be eventually sent to infinity (but 
in order to study the tree content 
of the model the limit $\kappa \to 0$ should 
be taken first). This model has not 
been studied yet:
\par
{\em Query}: What are the properties of the 
ensemble defined by (\ref{10}),
for reasonable choices of the correlation
inducing matrix $A$ ?  Even a study of the 
"semiclassical" limit alone would be of interest.

\section{Growing networks}
Recently, much activity has been devoted to 
the formulation of growing network
algorithms producing the so-called scale-free 
graphs (see, for example, refs. \cite{ba,gn,ke}). In these
models and at large "time" the average 
connectivity becomes stationary, except 
for the tail where finite size (time!) 
corrections are felt. A repeated use of 
such a growing network algorithm defines 
a statistical ensemble and, with this
strategy, it is not difficult to produce 
non-degenerate graphs only. The 
connectivity distribution cannot be chosen 
at will, it has a shape specific to
the model at hand. But one can usually 
adjust the parameters of the algorithm
to control the large vertex degree behavior. 
The major problem with this approach 
is that it is difficult to decide whether 
the results one obtains are generic
or just reflect the specific dynamics 
of a rather simple model. 
\par
Let me illustrate this point with 
an example in the next section.
\vspace{0.3cm}\par\noindent
\section{Graph diameters}
Consider {\em tree} graphs with 
connectivity distribution 
\begin{equation}
P_n = \frac{4}{n (n+1) (n+2)}
\label{11}
\end{equation}
They can be generated by the Barabasi-Albert 
growing network recipe \cite{ba,gn},
or by the algorithm presented in Sec. 3, 
provided the couplings are set to
\begin{figure}
\begin{center}
\epsfig{file=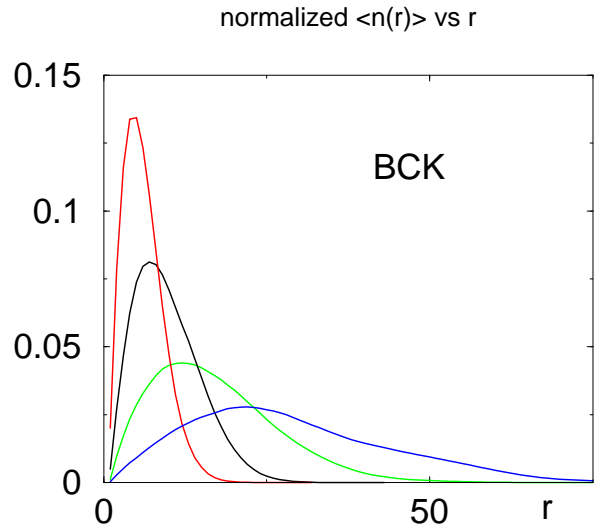, height=7cm,angle=0}
\end{center}   
\caption{\footnotesize{The normalized 
two-point function $\langle n(r) \rangle$
in the statistical ensemble defined in 
\cite{bck}  calculated for the     
number of nodes $N = 100, 400, 
1600, 6400$. The connectivity
distribution is given by eq (\ref{11}).}}  
\label{fig1}   
\end{figure}                           
\begin{figure}  
\begin{center}
\epsfig{file=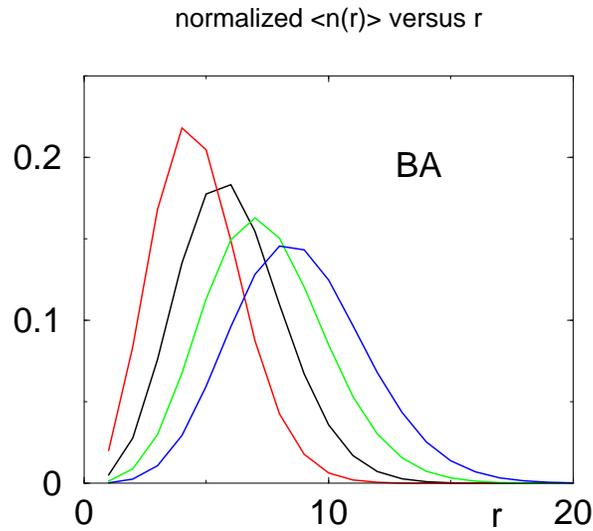, height=7cm,angle=0}
\end{center}        
\caption{\footnotesize{The normalized 
two-point function $\langle n(r) \rangle$
in the ensemble of graphs generated by 
the growing network algorithm     
proposed by Barabasi-Albert in \cite{ba}, 
calculated for the
number of nodes $N = 100, 400, 1600, 6400$. 
The connectivity
distribution is given by eq (\ref{11}).}}
\label{fig2}     
\end{figure}
$p_n = P_n/n$. For a given graph 
let $n(r)$ denote the number of 
vertices separated by geodesic 
distance $r$ from a randomly chosen
"reference" vertex. 
Averaging over the ensemble of graphs 
one is interested in and over the 
possible choices of the "reference" 
vertex, one gets a specific 
"two-point function" 
$\langle n(r) \rangle$, which can be used 
to define the average diameter of 
a graph. All this is easily done 
on a computer. The result, illustrated 
in Figs \ref{fig1} and 
\ref{fig2}, is that 
$\langle n(r) \rangle$ is very different in
the two models. In the 
Barabasi-Albert model the graph diameter 
grows like $\log{N}$, while in 
the model of Sec. 3 it grows like
a power of $N$ (see Fig. 
\ref{fig3}). Manifestly, the 
Barabasi-Albert model explores 
only a fraction of available phase-space. 
This is simply explained: the 
vertices of highest degree are 
the oldest ones and tend in this model to 
be close to each other. Consequently, 
the distance between other 
vertices is also much smaller than 
in a truly random tree. Another 
deviation from randomness in growing 
networks was observed earlier
by Callaway et al \cite{chkns}.
\begin{figure}     
\begin{center}
\epsfig{file=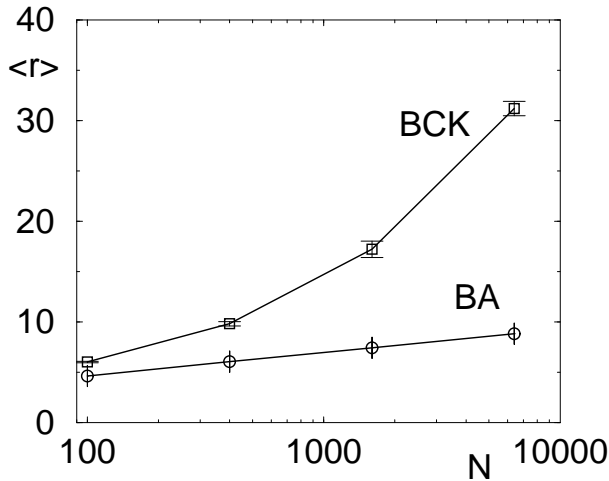, height=7cm,angle=0}
\end{center}     
\caption{\footnotesize{The average 
size of a graph $\langle r \rangle$ 
versus $N$ in the two models. It 
seen that in the Barabasi-Albert model the 
growth of $\langle r \rangle$ is logarithmic.}} 
\label{fig3}    
\end{figure}  
\par
\begin{figure}     
\begin{center}
\epsfig{file=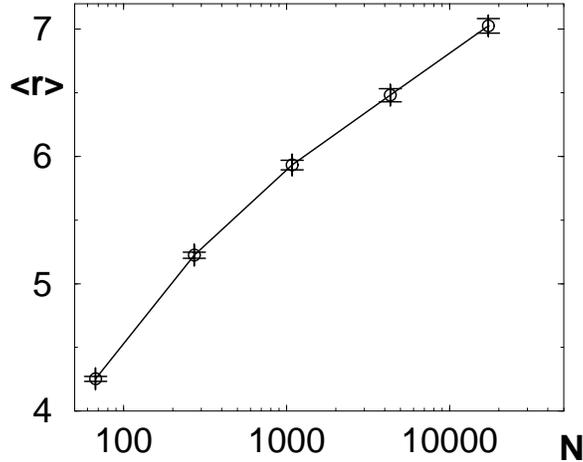, height=7cm,angle=0}
\end{center}     
\caption{\footnotesize{The average 
size $\langle r \rangle$  of the 
giant component 
versus $N = $average \#nodes in the 
component for general (degenerate) 
graphs with loops.  $N$ is not 
very large and finite size
correction to $\langle r \rangle$ 
is important. The connectivity 
distribution is (\ref{11}) for the 
full graph and falls also roughly 
like $\sim n^{-3}$ for the giant component.}}   
\label{fig4}  
\end{figure}

Incidentally, it appears that 
$\langle r \rangle \sim \log{N}$ in the 
ensemble of {\em degenerate} graphs 
with the same $P_n$ generated by the
algorithm of Sec. 3 (see Fig. \ref{fig4}).
Intuitively it is obvious that 
the growth of the diameter becomes slower 
when loops can be formed often 
enough since they produce "shortcuts".
\par 
Actually, the "small world" behavior 
$\langle r \rangle \sim \log{N}$ is found in
a large variety of networks. I do not 
know any rigorous derivation of this
result in a sufficiently general 
context. In the network community 
one often refers to ref. \cite{nsw}. 
Unfortunately, although ref. \cite{nsw} 
is otherwise an interesting paper, their derivation 
of this logarithmic growth is mathematically incorrect. They 
have in mind the Molloy-Reed construction, 
but they actually consider tree 
graphs with uncorrelated vertex degrees, 
and for such trees the diameter 
usually grows as a power of $N$.
\par
To see the mistake, notice that for a 
given "reference" vertex one has 
\begin{equation}
       1 + n(1) + n(2) + ... + n(r_{max}) = N      
\label{sum}
\end{equation}
\par\noindent
Newman et al replace all the quantities in 
(\ref{sum}) by their bulk average values. 
However, this is, in general, 
illegal. With each "reference" point 
is associated a specific sequence $n(1), 
n(2), ...\; $. The conditional probability that 
$n(r)=k$ differs from the bulk 
probability that a vertex has $k$ 
$r^{th}$-near-neighbors.
It depends on the sequence 
leading to $n(r)$. One has to attach 
probability measures to possible 
sequences in graphs and also to 
graphs. The problem is not trivial 
but it was solved by Ambj\o rn 
et al \cite{adj} precisely for the 
class of connected tree graphs 
considered in ref. \cite{nsw}. The 
result is that generically the 
Haussdorf dimension is finite and 
therefore the graph diameter
grows like a power of $N$, as 
already mentioned. Hence, I 
end this talk with another query:
\par
{\em Query}: What are the general 
conditions insuring that the "small
world" behavior $\langle r \rangle \sim \log{N}$ 
does actually hold as an
exact result for $N \to \infty$ ?
\par
A nice theorem awaits for being 
formulated and proved! 
\par
I wish to thank Serguei Dorogovtsev for pointing
out to me that an argument used in the original
version of this text is spurious.

\end{document}